\documentclass[fleqn,usenatbib,useAMS]{mnras}
\usepackage{newtxtext,newtxmath}

\usepackage{graphicx}	
\usepackage{amsmath}	
\usepackage{multicol}        
\usepackage{bm}		
\usepackage{pdflscape}	

\usepackage[normalem]{ulem} 
\usepackage[dvipsnames]{xcolor}
\usepackage{soul}

\usepackage{tikz}
\usetikzlibrary{calc, positioning}

\usepackage[percent]{overpic}

\usepackage{booktabs, multirow, makecell}

\setcellgapes{5pt}
\newcommand{\rh}{\, r_{\rm half}}
\newcommand{\rmax}{\, r_{\rm max}}
\newcommand{\msol}{\, M_{\odot}}

\newcommand{\pc}{\,{\rm pc}}
\newcommand{\kpc}{\,{\rm kpc}}

\newcommand{\mpc}{\,{\rm Mpc}}
\newcommand{\oversim}[2]{\protect{\mbox{\lower0.5ex\vbox{%
   \baselineskip=0pt\lineskip=0.2ex
   \ialign{$\mathsurround=0pt #1\hfil##\hfil$\crcr#2\crcr\sim\crcr}}}}} 
\newcommand{\bb}[1]{\ifmmode \mbox{\boldmath $ #1$} \else  \mbox{\boldmath $#1$} \fi}
\def\3{{\ss} }

\def\c12{{1\over 2}}

\def\d{{\rm d}}   
   
\def\plusplus{\raise 0.3ex\hbox{${\scriptstyle ++}$}{}}

\newcommand{\ra}{\rightarrow}
\def\and{{{\rm M}31}}

\def\gyr{\,{\rm Gyr}}

\def\vmax{\,v_{\rm max}}
\def\rmax{\,r_{\rm max}}
\def\smax{\,\sigma_{\rm max}}
\def\rh{\,r_{\rm half}}

\begin{document}

\title[The Heating Argument]{A dynamical attractor in the evolution of dwarf spheroidal galaxies}
\author[Pe\~narrubia \& Nadler]{Jorge Pe\~narrubia$^{1}$\thanks{Email: jorpega@roe.ac.uk}, Ethan O. Nadler$^{2}$\thanks{Email: enadler@ucsd.edu}\\
$^1$Institute for Astronomy, University of Edinburgh, Royal Observatory, Blackford Hill, Edinburgh EH9 3HJ, UK\\
$^2$Department of Astronomy \& Astrophysics, University of California, San Diego, La Jolla, CA 92093, USA}
\maketitle  

\begin{abstract}
We use controlled $N$-body experiments to study the dynamical evolution of dwarf spheroidal galaxies (dSphs) embedded in dark-matter (DM) haloes containing a large population of dark subhaloes. We show that stellar orbits subject to stochastic force fluctuations irreversibly gain energy and expand toward a dynamical attractor characterized by a stellar half-light radius $\rh \approx \rmax$ and a velocity dispersion $\sigma \approx 0.5\,v_{\rm max}$, where $v_{\rm max}$ is the peak circular velocity of the host halo at radius $r_{\rm max}$. This state is reached both in isolation and under tidal stripping, although tidal mass loss significantly accelerates the evolution.
Assuming that the Milky Way (MW) dSphs have reached this state, we find that the inferred halo masses collapse onto narrow sequences as a function of $\rh$. Under this assumption, MW satellites with $\rh \lesssim 1\,\mathrm{kpc}$ follow the tidal tracks of cuspy haloes, while larger systems deviate in a manner consistent with cored DM profiles. Moreover, the mass--luminosity relation follows the slope expected from abundance matching, but with halo masses systematically lowered from their peak values at fixed luminosity.
These results suggest that the structural diversity of dSphs is largely an evolutionary outcome driven by internal heating and tides, rather than by the conditions of star formation. This framework predicts that isolated, early-quenched dSphs should have systematically larger sizes than satellites, a prediction testable with upcoming surveys.
\end{abstract}
  
\begin{keywords}
Galaxy: kinematics and dynamics; galaxies: evolution; Cosmology: dark matter.
\end{keywords}

\section{Introduction}\label{sec:intro}
Dwarf spheroidal galaxies (dSphs) are the faintest and most dark matter (DM)-dominated galaxies in the Universe, and as such play a fundamental role in galaxy formation as well as in on-going efforts to understand the nature of DM (e.g. Mateo 1999; Gilmore et al. 2007; Simon 2019).

Traditionally, attempts to infer the amount and distribution of DM in dSphs using stellar kinematics have relied on the Jeans equations, which suffer from the well-known mass–anisotropy degeneracy (e.g., Read et al. 2021). This degeneracy is minimized at a radius $\lambda \rh$, where $\lambda \simeq 1.8$ and $\rh$ is the stellar half-light radius (Errani et al. 2018). From the virial theorem it follows that the mass enclosed scales as $M_h(<\lambda \,\rh) = \mu_0 \,\lambda \,\rh \sigma^2 / G$, where $\sigma$ is the luminosity-averaged velocity dispersion. The proportionality factor $\mu_0$ depends on the spatial segregation of stars within the DM halo, $\rh/\rmax$, on the shape of the stellar density profile, and also on the unknown halo potential (Errani et al. 2018; Splawska et al. 2026).

However, measuring $M_h(<\lambda\, \rh)$ alone does not constrain the total DM mass, $M_h$, because the distribution of stellar tracers within the halo is unknown. In particular, a fundamental degeneracy exists between $M_h$ and the ratio $\rh/\rmax$ (e.g., Pe\~narrubia et al. 2008a). Applying simple virial mass estimators to Milky Way (MW) dSphs—with a fixed $\mu_0$ coefficient—reveals a narrow scaling relation between size and enclosed DM, approximately consistent with all MW dSphs residing in cuspy haloes of $M_h\sim10^9\,\msol$, with varying degrees of stellar segregation (Walker et al. 2009). Meanwhile, abundance-matching analyses generally assign the faintest observed dSphs to lower-mass haloes, $\sim10^8\,\msol$, but these studies do not directly leverage dynamical information (Jethwa et al. 2018; Nadler et al. 2020).

Proper motion measurements can be incorporated into Jeans modelling to break the mass–anisotropy degeneracy for individual galaxies. However, in flattened dSphs, the inferred halo potential also depends on the unknown 3D orientation of the stellar distribution along the line of sight, introducing additional uncertainty (Vitral et al. 2025b). Overall, the total halo masses and density profiles of individual dSphs remain uncertain, limiting our understanding of both the faint end of the galaxy–halo connection and the microphysical nature of DM (Nadler et al. 2020, 2021).
We are therefore motivated to explore alternative ways to break these degeneracies.


Here, we build on the results of Pe\~narrubia et al. (2025, hereafter P25), who showed that stellar orbits in DM haloes that contain substructures heat up and expand due to repeated encounters. This expansion is generic to tracers experiencing stochastic force fluctuations. For example, similar heating occurs in fuzzy-DM haloes (Dutta Chowdhury et al. 2023) and in galaxies subject to recurrent supernova-driven gas outflows (Hashim et al. 2023). Subhalo-induced fluctuations generate subtle corrugations in the surface brightness maps of dSphs that are detectable with deep photometric data (Vitral et al. 2025a).
Crucially, P25 found that as the stellar half-light radius approaches the halo peak velocity radius, $\rmax$, the expansion slows and the velocity dispersion reaches a maximum value, $\sigma \simeq 0.54\,\vmax$, suggesting that these systems evolve toward a dynamical attractor. 

In P25, dSphs were modelled as isolated systems embedded in cuspy DM haloes that contain a sizeable population of subhaloes in dynamical equilibrium within a static potential. This paper extends that work to (1) explore how different DM halo profiles affect the gravothermal expansion of stellar tracers, (2) determine the impact of tidal stripping, and (3) investigate the implications of dynamical heating for observed MW dSph satellites.

The paper is organized as follows: In \S\ref{sec:virial} we present numerical experiments following the expansion of stellar orbits toward the dynamical attractor in haloes with dark substructures; in \S\ref{sec:HA} we apply the virial equations to a catalogue of MW dSphs assuming they have reached this attractor; \S\ref{sec:dis} discusses the implications of our results.

\begin{figure*}
\begin{center}
\includegraphics[width=155mm]{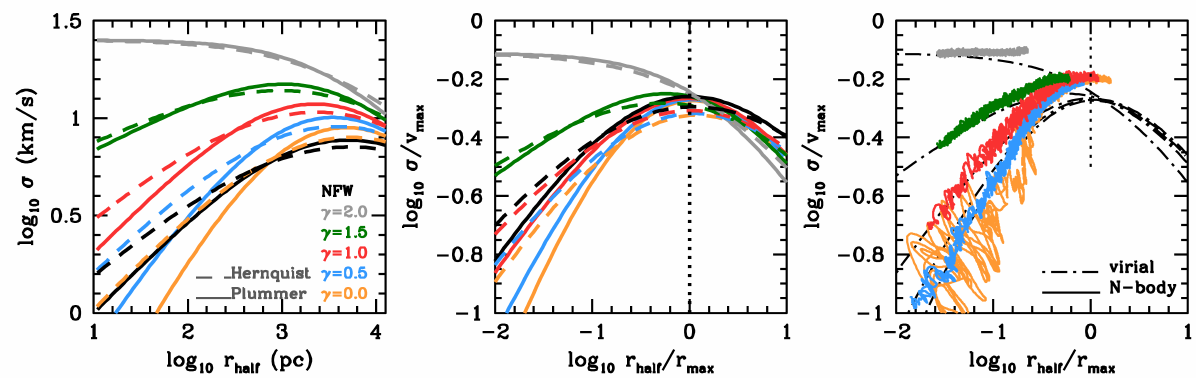}
\end{center}
\caption{
{\it Left panel:} Stellar velocity dispersion $\sigma$ as a function of half-light radius $\rh$, derived from the virial theorem for stellar tracers with Plummer (solid) or Hernquist (dashed) profiles embedded in Dehnen (colored) and NFW (black) haloes with $M_h=10^9\,\msol$ and $c_h=2.26\,\kpc$. The dependence on the stellar profile is weak in cuspy haloes and increases as the inner density slope approaches $\gamma\to0$.
{\it Middle panel:} Same relations normalized to the halo peak velocity $\vmax$ and radius $\rmax$. As $\rh \ra r_{\rm max}$ (vertical dotted line), the velocity dispersion peaks at $\eta_\sigma=\sigma_{\rm max}/\vmax\simeq0.54$ for a Plummer profile and $0.48$ for a Hernquist profile.
{\it Right panel:} Comparison with $N$-body models (see text). Stochastic heating drives progressive expansion of the tracers, proceeding more rapidly in haloes with shallower density profiles. At early times, the evolution follows the virial-theorem prediction (dash–dotted lines). In all cases, the stellar velocity dispersion peaks at $\eta_\sigma\simeq0.63$, independent of the halo profile, exceeding the virial prediction by a factor $\simeq1.16$.
}
\label{fig:rhsig}
\end{figure*}

\section{Virial equations}\label{sec:virial}
\subsection{Stellar kinematics}\label{sec:theorem}
In equilibrium, the luminosity-averaged 1D velocity dispersion of tracer stars with a 3D luminosity profile $\nu(r)$ embedded in an extended DM halo with a spherical mass profile $M_h(<r)$ can be computed from the virial theorem as $\sigma^2=-W$, where 
\begin{align}\label{eq:W}
  W=-\frac{4\pi }{3}\int_0^\infty \d r\, r^2\,\nu(r) \frac{G M_h(<r)}{r},
\end{align}
is the potential energy of stellar particles moving in the DM halo (e.g. Errani et al. 2018). 
Since dSphs are DM-dominated objects, in what follows we assume that stars behave like mass-less tracers with a  Plummer (1905) or Hernquist (1990) profiles normalized such that $\int \d^3r \,\nu(r)=1$.

To gain physical insight, let us model the DM halo analytically using a Dehnen (1993) mass profile
\begin{align}\label{eq:mdehn}
  M_h(<r)= M_h \, \bigg(\frac{r}{r+c_h}\bigg)^{3-\gamma},
\end{align}
where $\gamma\ge 0$. 
The density profile associated with~(\ref{eq:mdehn}) scales as $\rho_h \sim r^{-\gamma}$ at small radii $r\ll c_h$ and as $\rho_h \sim r^{-4}$ at large radii $r\gg c_h$. Note that this halo model has a finite total mass, $M_h(<r)\to M_h$ as $r\to \infty$. 
Of particular cosmological interest are the `cuspy' profile with $\gamma=1$ (Hernquist 1990), as this matches the universal profile found in CDM simulations of structure formation (e.g. Navarro, Frenk \& White 1996), and the `cored' profile with $\gamma=0$.
From Equation~(\ref{eq:mdehn}), it is trivial to derive the circular velocity of the halo, $v_c(r)=\sqrt{GM_h(<r)/r}$, which peaks at $\rmax=(2-\gamma)\,c_h$. Notice that $\rmax$ does not exist for haloes with $\gamma> 2$, as their circular velocity profiles decrease with radius monotonically. 

One can use Equation~(\ref{eq:W}) to study how the stellar velocity dispersion depends on the segregation of stars within the DM halo.
The left panel of Fig.~\ref{fig:rhsig} shows the velocity dispersion computed numerically from Equation~(\ref{eq:W}) for stellar tracers with Plummer (solid lines) and Hernquist (dashed lanes) profiles embedded in an NFW potential (black lines) and in Dehnen potentials with a wide range of inner density slopes ($\gamma$). In these models, we set the total halo mass\footnote{In the case of the NFW profile, we use a virial mass $M_{\rm vir}=M_h$ with a concentration $r_{\rm vir}/r_s=20$.} to $M_h=10^9\msol$ and the scale radius to $c_h=2.26\kpc$.
Independently of the shape of the stellar mass and distributions, we find that the stellar velocity dispersion grows as a function of $\rh$, and that it becomes roughly constant at large radii. Differences between the Plummer and Hernquist stellar profiles are mostly visible for models with small half-light radii.

These differences are highlighted by normalizing the stellar velocity dispersion by $\vmax$ and the half-light radius by $\rmax$ (middle panel), which shows that for models with $\gamma<2$ the stellar velocity dispersion peaks roughly at the same radius as the halo circular velocity (marked with a vertical dotted line for reference). We also notice that the ratio $\smax/\vmax$ barely depends on the choice of DM profile, and only slightly on the stellar profile. More specifically, for stars following a Plummer profile $\smax/\vmax\simeq 0.54$, whereas for those with a Hernquist profile $\smax/\vmax\simeq 0.48$ for $0.3\lesssim \rh/\rmax\lesssim 3$. 

As shown in P25, stellar orbits embedded in a halo populated by DM substructures undergo gradual heating due to repeated encounters, causing the system to expand while remaining close to virial equilibrium. Following the methodology of P25, we construct a suite of 
$N$-body experiments that track the dynamical evolution of stellar tracers inside haloes hosting a substantial population of subhaloes with a mass function rescaled from the Aquarius simulation (Springel et al. 2008). The orbits of the subhaloes are sampled from a radially anisotropic Osipkov-Merritt distribution function (see P25 for full details). At $t=0$, the stellar component follows a Plummer profile with a half-light radius $\rh=30\pc$ in dynamical equilibrium within the DM potential. Stellar and subhalo orbits are integrated simultaneously until $t=5000\,\omega^{-1}$, where $\omega=\sigma/\rh$ is the orbital frequency measured at $t=0$. For simplicity, this set up ignores the effects of subhaloes on the smooth host potential (see P25).

Repeated encounters with subhaloes inject kinetic energy into stellar orbits, driving a gradual gravothermal expansion within the (static) DM potential. The response depends on the inner slope of the halo: in models with $\gamma<2$, the growth of $\rh$ is accompanied by a rise in $\sigma$, whereas in the isothermal model ($\gamma=2$) the system expands at nearly constant velocity dispersion. As the half-light radius approaches the peak-velocity radius, $\rh\gtrsim 0.3\rmax$, the velocity dispersion starts to plateau. In isolation, at maximum expansion we measure $\smax/\vmax\simeq 0.63$ independent of the halo profile, which is slightly higher than predicted by the virial theorem. This offset may reflect the breakdown of equilibrium once $\rh\gtrsim \rmax$ (see \S\ref{sec:evol}).

The efficiency of gravothermal expansion also varies strongly with halo structure: it proceeds slowly in isothermal cusps ($\gamma=2$), and becomes progressively more efficient as the density profile flattens. Notably, core-dominated haloes exhibit large fluctuations in stellar velocity dispersion up to a factor of $\approx 3$ at a fixed $\rh$. This may be due to inefficient phase-space mixing in cored haloes (e.g. Errani et al. 2025), which allows non-equilibrium features induced by subhalo perturbations to persist for extended periods.

\subsection{The Heating Argument}\label{sec:total}
As shown in the previous Section, stellar tracers that have fully expanded in the DM halo ($\rh/\rmax\sim 1$) have a fixed ratio $\smax/\vmax\approx 0.5$--$0.6$, with scant sensitivity to the shape of the stellar or halo profiles. This can be used to derive the total halo mass from the stellar half-light radius and average velocity dispersion of dSphs. For Dehnen haloes, the derivation is analytical. The condition $\d v_c/\d r=0$ happens at $\rmax=(2-\gamma)\,c_h$. Hence, $\vmax^2=v_c^2(\rmax)=(2-\gamma)^{2-\gamma}/(3-\gamma)^{3-\gamma}(GM_h/c_h)$. Defining $\eta_r\equiv\rh/\rmax$ and $\eta_\sigma\equiv \sigma/\vmax$, and re-arranging yields 

\begin{align}\label{eq:totalm}
M_h(\eta_r,\eta_{\sigma})=\bigg(\frac{3-\gamma}{2-\gamma}\bigg)^{3-\gamma}\frac{1}{\eta_r\eta_\sigma^2}\frac{\rh\,\sigma^2}{G}.
\end{align}
The dependence of the mass estimator in Equation~(\ref{eq:totalm}) on the inner slope is mild; in particular, the factor $f_\gamma=[(3-\gamma)/(2-\gamma)]^{3-\gamma}$ varies from $f_0\simeq 3.375$ to $f_1=4.0$. Furthermore, according to Fig.~\ref{fig:rhsig}, the peak velocity dispersion derived from the virial theorem is approximately $\eta_\sigma\sim 0.5$ for half-light radii in the range $0.3\lesssim \eta_r\lesssim 3$. This value holds for inner slopes $\gamma<2$, and appears to be barely sensitive to the choice of stellar profile.


In isolation, dSphs expand progressively until they reach the peak velocity radius of their host DM halo, $\eta_r\sim 1$, as shown in Fig.~\ref{fig:rhsig}. E.g., the fiducial dSph models with $\gamma=1$ plotted in Fig.~\ref{fig:rhsig} must evolve $\sim 5000$ crossing times, which corresponds to $\sim 14\gyr$, to reach the peak velocity radius. In practice, some galaxies may not reach this point of their evolution within a Hubble time, whereas others may grow beyond the peak velocity radius if heating continues uninterruptedly. 


Below, we show that tidal stripping combined with stochastic heating act together to accelerate the dynamical evolution of stellar tracers in DM haloes toward $\eta_r\to 1$ and $\eta_{\sigma}\to 0.5$. This convergence forms the basis of what we term the {\it heating argument} (HA). We therefore define masses in the HA limit as $M_{\mathrm{HA}}=M_h(\eta_r=1,\eta_{\sigma}=0.5)$ based on Equations~(\ref{eq:totalm}).

\begin{figure}
\begin{center}
\includegraphics[width=75mm]{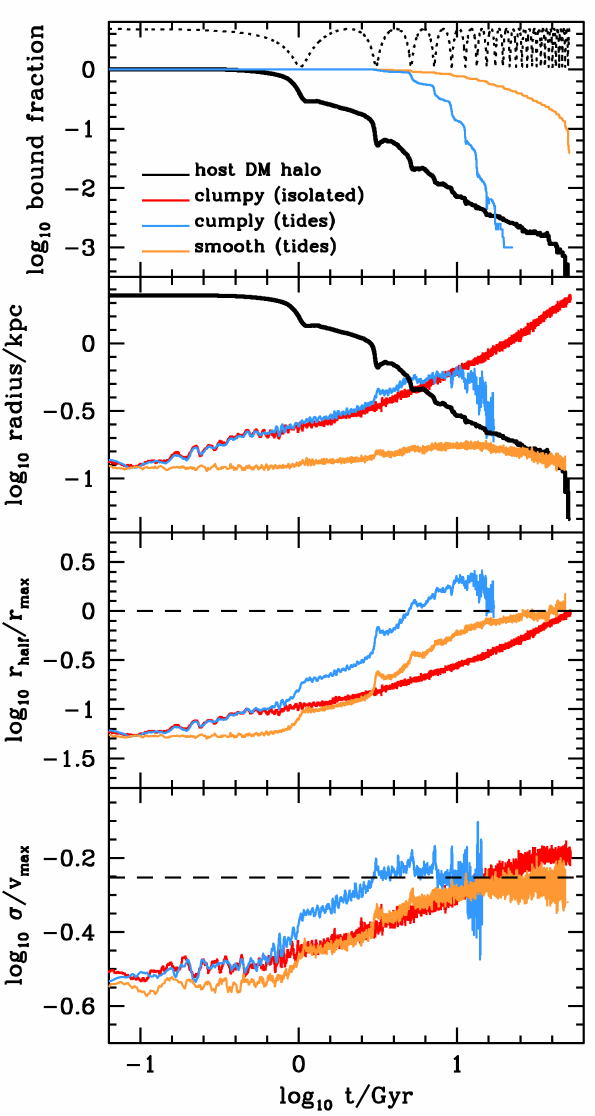}
\end{center}
\caption{{\it Top panel:} Time evolution of the bound subhalo mass fraction (black solid line) for a dSph with $M_h=10^9\,\msol$ and $c_h=2.26\,\kpc$ orbiting a MW-like halo with $M_G=10^{12}\,\msol$ and $c_G=21.5\,\kpc$. The orbit has apocentre $r_{\rm apo}=120\,\kpc$ and pericentre $r_{\rm peri}=25\,\kpc$ (shown with a black dotted line in arbitrary units). The bound stellar mass fraction of smooth and clumpy halo models are shown in orange and blue, respectively. Notice that while the extended DM halo initially shields the stars from stripping, subhaloes enhance stellar mass loss at late times.
{\it Second panel:} Evolution of the DM peak velocity radius (black solid line), which follows the tidal track $r_{\rm max}/r_{\rm max,0}=(M/M_0)^\kappa$ with $\kappa=0.44$ (Errani \& Pe\~narrubia 2020). Stellar half-light radii for the smooth and clumpy models are shown in orange and blue; the red line shows a clumpy halo evolving in isolation. 
{\it Third panel:} Stellar segregation within the halo, $\eta_r=\rh/r_{\rm max}$, as a function of time. All models evolve toward $\eta_r\to1$ (black dashed line), with the fastest convergence occurring in tidally stripped, clumpy haloes.
{\it Bottom panel:} Stellar-to-DM velocity dispersion ratio, $\eta_\sigma=\sigma/v_{\rm max}$; the asymptotic limit $\eta_\sigma\to0.5$ is indicated by the black dashed line.
}
\label{fig:evol}
\end{figure}

\subsection{Evolution in a tidal field}\label{sec:evol}
The previous section considered dynamical models of isolated dSphs. However, since most dSphs orbit within larger host haloes such as the Milky Way, it is necessary to assess how their internal dynamics evolve after accretion. We therefore repeat the above experiments by placing our dSph models on an eccentric orbit within a MW-like host, represented by a static spherical Hernquist (1990) potential with total mass $M_G = 10^{12},\msol$ and scale radius $c_G = 21.5,\kpc$.

In this set-up, the dSph moves within the fixed MW potential, while stars and subhaloes, initially bound to the dSph, evolve under the combined gravitational field of the MW and the dSph halo. In addition, stars experience the individual gravitational forces generated by each individual subhalo. Accordingly, we simultaneously solve three coupled sets of equations describing: (i) the orbit of the dSph in the MW potential; (ii) the motion of subhaloes in the combined MW+dSph potential; and (iii) the motion of stars in the MW+dSph potential, including the additional contribution from individual subhaloes (see Appendix~C of Pe\~narrubia et al. 2023 for details on the numerical integration). 
For illustration, the orbit of the dSph about the MW has an apocentre $r_{\rm apo} = 120\,\kpc$ and pericentre $r_{\rm peri} = 25\,\kpc$. Varying these parameters does not affect our qualitative conclusions.
This semi-analytic set-up assumes that tidal stripping preserves the functional form of the halo density profile, allowing only the bound halo mass and scale radius to evolve with time (see \S\ref{sec:dis} for discussion). To define bound/unbound we track the binding energy of individual subhalo and stellar orbits in the dSph potential. At each time step, we record the fraction of subhaloes lost to tides. Since subhaloes and the dSph halo share the same distribution function by construction, this is also the mass loss fraction of the dSph. 
To account for the structural evolution of the dSph halo induced by mass loss, we adopt the tidal tracks identified by Errani \& Pe\~narrubia (2020), who found that cuspy haloes subject to stripping evolve according to $c_h/c_{h,0} = (M_h/M_{h,0})^\kappa$, with $\kappa \simeq 0.415$ (see their Fig.~5). We set the initial dSph halo parameters to the fiducial halo model of P25, i.e. $M_{h,0} = 10^9\,\msol$ and $c_{h,0} = 2.26\,\mathrm{kpc}$. We have independently verified that self-consistent $N$-body simulations are in good agreement with this parametrized description of tidal evolution.

The upper panel of Fig.~\ref{fig:evol} shows the evolution of the bound mass fraction of dark matter and stars in models without subhaloes and with subhaloes (black, orange, and blue lines, respectively). 
As expected, the dSph halo mass is approximately constant until the dwarf galaxy crosses the first pericentre at $t\simeq 1\gyr$ (for reference, the orbital radius evolution with a dotted line in arbitrary units). From this time on, the bound mass drops at each subsequent pericentric passage.

Initially, the stellar component is deeply embedded within the DM halo, with $r_{\rm half,0}=120\pc$, corresponding to $\rh/\rmax\simeq 0.05$. This strong spatial segregation shields the stars from external tidal forces (e.g. Pe\~narrubia et al. 2008b). Indeed, this model must lose more than $\sim 99\%$ of its initial halo mass and complete roughly two orbital periods before significant stellar mass loss occurs. Crucially, stellar stripping is more efficient in the clumpy halo (blue line) than in the smooth halo (orange line). In the presence of dark subhaloes, stars begin to be tidally stripped earlier and also experience a higher mass loss rate. 

The second panel from the top shows that the halo peak-velocity radius $\rmax$ (black solid line) decreases significantly due to tidal mass loss. Simultaneously, the half-light radius of stars in the clumpy halo (blue line) grows driven by stochastic heating induced by subhalo perturbations. This expansion lowers the mean binding energy of stellar orbits, thus enhancing stellar mass loss. In contrast, stars orbiting in a smooth halo (orange line) experience little heating, which prolongs the time they remain bound to the dSph potential.

A comparison with the isolated models (red line) shows that the expansion of the stellar component is barely affected by external tides until the stellar half-light radius becomes comparable to the halo peak-velocity radius, $\rh\simeq\rmax$, at $t\simeq3\gyr$. Once this threshold is reached, tidal stripping leads to a contraction of $\rh$. At this point, the galaxy has lost most of the stellar particles to tides.

The third and fourth panels show the evolution of $\eta_r=\rh/\rmax$ and $\eta_\sigma=\smax/\vmax$, respectively. All models evolve toward $\eta_r\to 1$ and $\eta_\sigma\to 0.5$ (horizontal dashed lines). Convergence is fastest for dSphs evolving in a tidal field within a clumpy halo (blue), while tidally stripped smooth models (orange) also approach the attractor more rapidly than the isolated clumpy case (red).
These results agree with self-consistent $N$-body simulations of tidally limited dSphs, which show that once $\rh\approx\rmax$, the stellar velocity dispersion satisfies $\sigma\approx0.5\vmax$ largely independent of the initial stellar segregation (see Fig.~13 of Errani et al. 2022). 
We note that in isolated models $\eta_\sigma$ is a factor $\sim1.16$ higher than in tidally stripped systems. This offset is driven by stellar particles scattered onto weakly bound, large-apocentre orbits that do not efficiently phase-mix. These stars inflate the global velocity dispersion and maintain the stellar component in a persistently super-virial state ($2K>|W|$), as shown in Fig.~\ref{fig:rhsig}. In a tidal field, such loosely bound stars are preferentially removed, leaving a bound remnant that converges toward $\eta_\sigma\simeq0.5$, close to the virial expectation and to the value found in self-consistent $N$-body simulations. 

Thus, both external (tidal) and internal (heating) processes drive $\rh/\rmax$ and $\sigma/v_{\max}$ toward a common dynamical attractor that reflects the depth of the halo potential. These results suggest that many MW dSphs may have reached the HA attractor by the present day; we now demonstrate the implications of this argument for key dSph scaling relations.

\section{Application to the MW dSphs }\label{sec:HA}
Here, we apply the HA to the catalogue of half-light radii and luminosity-averaged velocity dispersions of MW dwarf galaxies published by Pace et al. (2022). Statistical uncertainties are estimated assuming that the quoted observational errors are Gaussian.

\begin{figure}
\begin{center}
\includegraphics[width=84mm]{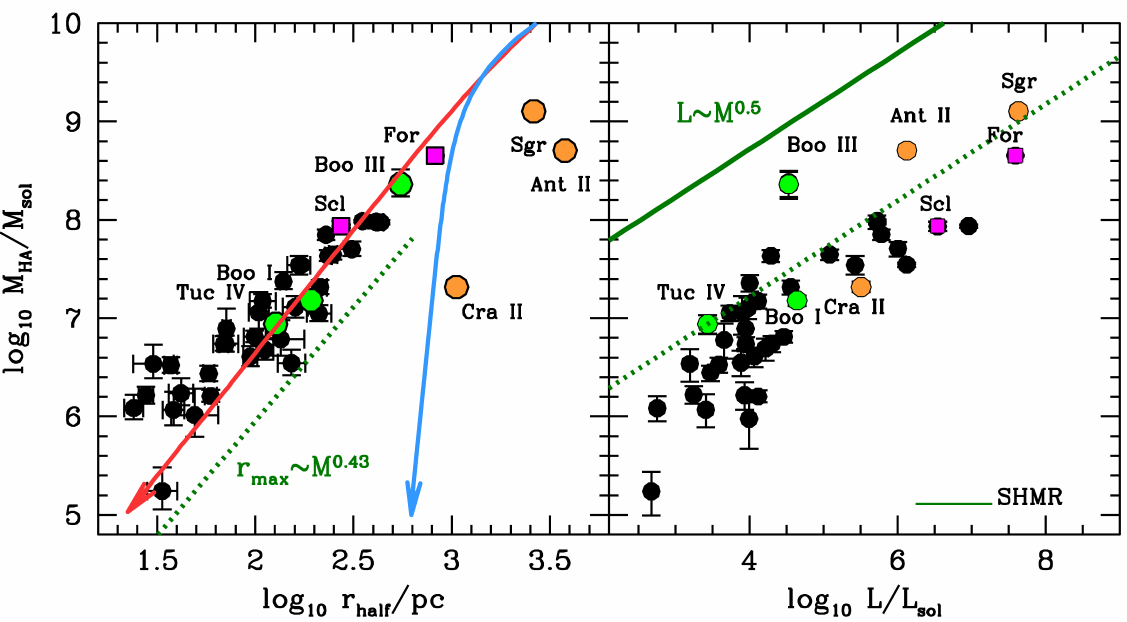}
\end{center}
\caption{{\it Left panel:} Dwarf galaxy masses inferred from Equation~(\ref{eq:totalm}), with $(\eta_r,\eta_\sigma)=(1,0.5)$, as a function of half-light radius. Here, we adopt $\gamma=1$, although $M_{\rm HA}$ values are largely insensitive to this choice. Orange circles indicate galaxies with associated stellar streams, while green circles mark systems exhibiting signatures of tidal stripping (Pace et al. 2022). Magenta squares correspond to galaxies that may reside in cored DM haloes. Red and blue curves show the tidal tracks expected for galaxies embedded in cuspy ($\gamma=1)$ and cored ($\gamma=0$) haloes, respectively (see text). Note that dSphs with $\rh\lesssim 1\kpc$ lie close to the cuspy tidal track, as well as to the slope of the subhalo mass--size relation in the Aquarius CDM simulation (shown with a green dotted line with arbitrary normalization), $r_{\rm max}\sim M^{0.43}$ (Springel et al. 2008). {\it Right panel:} HA masses as a function of dSph luminosity. The green-solid line shows the best-fit luminosity--peak halo mass relation derived from the MW satellite luminosity function (Nadler et al.\ 2020). The green-dotted line shows the same relation with halo masses lowered by 1.5 dex at a fixed luminosity.}
\label{fig:obs}
\end{figure}

The upper panels of Fig.~\ref{fig:obs} show the halo masses inferred from Equation~(\ref{eq:totalm}) under the HA expectation $(\eta_r,\eta_\sigma)=(1,0.5)$. We adopt a cuspy halo profile, $\gamma=1$, although this choice has only a modest effect on the inferred masses; for example, assuming a cored halo with $\gamma=0$ would lower the mass estimates by only a factor of $\sim 0.84$. The most striking feature of the mass--size relation in the upper-left panel of Fig.~\ref{fig:obs} is the clear separation between galaxies with $\rh\lesssim 1\kpc$ and those with $\rh\gtrsim 1\kpc$. All dSphs in the former group lie along a relatively narrow sequence that runs parallel to the mass--size relation of subhaloes in the Aquarius CDM simulation, $r_{\rm max}\sim M^{0.43}$ (Springel et al. 2008).
Interestingly, this sequence also runs approximately parallel to the tidal track of cuspy haloes, $g(x)=2^\mu x^\eta/(1+x)^\mu$ (red line), where $x=M_h/M_h(t=0)$, $g=\rmax/\rmax(t=0)$, and $(\mu,\eta)=(-0.3,0.4)$ (Pe\~narrubia et al. 2010). In contrast, the tidal track for cored haloes, $(\mu,\eta)=(-1.3,0.05)$ (blue line) deviates from the observed trend. 

It is important to emphasize that the initial point of each tidal track ($x=1$) is arbitrary in this exercise.  Fig.~\ref{fig:obs} chooses the starting point so that the cuspy tidal track passes through the locus defined by galaxies with $\rh\lesssim 1\kpc$. This choice helps to visualize that galaxies with $\rh\gtrsim 1\kpc$ lie systematically below the extrapolation of the scaling relation traced by the smaller dSphs. Given that all of the dSphs that lie below the cuspy track (i.e., Crater~2, Antlia~2, and Sagittarius) exhibit associated tidal streams, one possible interpretation is that these larger systems are embedded in cored DM haloes (Borukhovetskaya et al.\ 2022; Zhang et al.\ 2024).

Pace et al. (2022) also identify three dSphs that may be losing mass to tides (Boo~I, Boo~III, and Tuc~IV, shown in green). These systems lie close to the main scaling relation. This may be because they are evolving along a cuspy tidal track, or because tidal effects have not yet significantly altered their internal structure. Likewise, although several dynamical studies suggest that Fornax and Sculptor reside in cored dark matter haloes (e.g. Boldrini et al. 2021), neither galaxy shows a noticeable departure from the scaling relation, suggesting that even if they reside in cored haloes, they are unlikely to be undergoing substantial stellar mass loss at present. This is consistent with their relatively large pericentric distances, $r_{\rm peri}\sim50~\kpc$ (Sculptor) and $\sim80~\kpc$ (Fornax) (Battaglia et al. 2022; Pace et al. 2022).


The right panel of Fig.~\ref{fig:obs} shows the relation between dSph masses derived from the HA and their luminosities. We infer that dSphs lie along a scaling relation $L \sim M_{\mathrm{HA}}^{0.5}$ with relatively small scatter, despite the fact that luminosity is an independent observable that need not correlate with quantities derived from the HA. Remarkably, the $L$--$M_{\mathrm{HA}}$ relation follows the same slope as the best-fit stellar-to-peak halo mass relation (SHMR) derived by applying abundance matching to MW satellites (Nadler et al.\ 2020). More specifically, the observed relation is consistent with the SHMR if peak halo masses are lowered by $\sim 1.5~\mathrm{dex}$ at fixed luminosity\footnote{For galaxies with associated tidal streams, both their mass {\it and} luminosity were higher at the time of accretion.}, supporting a picture in which the DM haloes of dSphs are substantially stripped after infall. The small scatter in the inferred $M_{\mathrm{HA}}$--$L$ relation is remarkable given that different subhaloes should undergo different amounts of stripping depending on their orbits and infall times; a dedicated study is needed to understand how a relatively small SHMR scatter is preserved.


\section{Discussion \& Summary}\label{sec:dis}
In dSphs where the gravitational potential exhibits stochastic fluctuations, stellar orbits tend to gain energy and expand toward a dynamical attractor in which $\eta_r=\rh/\rmax\approx 1$ and $\eta_\sigma=\sigma/\vmax \approx 0.5$.
This process does not depend on the physical origin of the fluctuations: the same convergent behaviour arises in haloes populated by substructure, in fuzzy-DM haloes where wave interference produces density granularity, and in galaxies subject to recurrent supernova-driven gas outflows.
Figs.~\ref{fig:rhsig} and~\ref{fig:evol} demonstrate that this state is reached both in isolation and under tidal stripping. The attractor thus represents a generic evolutionary outcome rather than the imprint of star formation in the progenitor DM halo. %

Quantities estimated under the {\it heating argument} (HA) assume that dSphs have reached this state.
In practice, it is not possible to determine whether individual dSphs have fully converged to the attractor, since their global dark matter halo potentials are generally unknown. Because tidal stripping accelerates convergence, early-accreted dSphs are expected to lie closer to the attractor\footnote{In dSphs with multiple chemodynamical populations,
stars with different ages may lie at different points along their evolution toward the attractor.}   
However, for systems still expanding the values of $M_{\rm HA}$ derived from Equation~(\ref{eq:totalm}) systematically underestimate the true present-day halo masses, since $\eta_r $ and $\eta_\sigma $ remain below their asymptotic values. 
In this context, we emphasize that Equation~(\ref{eq:totalm}) assumes equilibrium haloes with a Dehnen (1993) profile, whereas tidal stripping can bring galaxies out of equilibrium and modify the outer structure of dark matter haloes (Errani \& Navarro 2021). 
Thus, further work is needed to calibrate the relation between $M_{\rm HA}$ and the bound masses of dSphs.

Remarkably, when evaluated under the HA, the masses of MW dSphs collapse onto narrow, well-defined sequences as functions of their half-light radii and luminosities. In particular, dSphs with $\rh \lesssim 1,\mathrm{kpc}$ follow a scaling relation parallel to the tidal track of cuspy DM haloes, indicating that their structural diversity can be understood as an evolutionary outcome driven by internal heating and tidal mass loss. In contrast, dSphs with $\rh \gtrsim 1,\mathrm{kpc}$ deviate from this track and have properties consistent with cored halo profiles. 
In this context, dSphs with large DM cores moving in a tidal field systematically evolve away from the observed sequences. The relatively small scatter of the $M_{\mathrm{HA}}$--$\rh$ relation of galaxies with $\rh\lesssim 1\kpc$ may potentially place tension on DM models that generically produce cores in low-mass haloes, such as fuzzy and self-interacting DM (Hu et al. 2017; Tulin \& Yu 2018).
Despite luminosity being independent of the quantities entering the HA, we also find a tight relation $L \propto M_{\mathrm{HA}}^{0.5}$ with small scatter. The slope matches that of the stellar–halo mass relation inferred from abundance matching, but is shifted toward lower halo masses by $\sim1.5$ dex, consistent with substantial tidal stripping of dSph haloes after infall.


Our models show that, once stellar stripping begins, the expansion of dSphs halts and reverses, while the velocity dispersion decreases. Consequently, dSphs that form with identical internal structure and age in haloes with the same mass are expected to be more extended and dynamically hotter in the field than when acted on by tidal stripping. This prediction may be testable with forthcoming surveys of quenched satellite and isolated dSphs throughout the Local Volume (for example, LSST is expected to detect dwarf galaxies as faint as $M_V\sim -5$ out to $\sim 1$--$2\mpc$, e.g. Mutlu-Pakdil et al. 2021). 

\section*{Acknowledgements}
The authors would like to thank the organizers of the meeting ``Small-Scale Structure of the Universe and Self-Interacting Dark Matter''  in Valencia, Spain, where part of this work was performed. We are grateful to the anonymous referee, Tara Dacunha, Rapha{\"{e}}l Errani, Frank van den Bosch, Eduardo Vitral, and Matthew Walker for comments on the manuscript.

\section*{Data availability}
No data were generated for this study.

{}

\end{document}